\documentclass[twocolumn]{aastex63}

\usepackage{color}

\submitjournal{ApJ}

\shorttitle{Screening Effects on EC rates and SN Ia Nucleosynthesis}
\shortauthors{Mori et al.}

\begin{document}

\title{Screening Effects on Electron Capture Rates and Type Ia Supernova Nucleosynthesis}

\author[0000-0003-2595-1657]{Kanji Mori}
\altaffiliation{Research Fellow of Japan Society for the Promotion of Science}
\affiliation{Graduate School of Science, The University of Tokyo,
7-3-1 Hongo, Bunkyo-ku, Tokyo, 113-0033 Japan}
\affiliation{Division of Science, National Astronomical Observatory of Japan, 2-21-1 Osawa, Mitaka, Tokyo 181-8588, Japan}

\author[0000-0002-5500-539X]{Toshio Suzuki}
\affiliation{Department of Physics, College of Humanities and Sciences, Nihon University, 3-25-40 Sakurajosui, Setagaya-ku, Tokyo 156-8550, Japan}
\affiliation{Division of Science, National Astronomical Observatory of Japan, 2-21-1 Osawa, Mitaka, Tokyo 181-8588, Japan}

\author{Michio Honma}
\affiliation{Center for Mathematical Sciences, University of Aizu, Tsuruga, Ikki-machi, Aizu-Wakamatsu, Fukushima 965-8580, Japan}
\author[0000-0003-2305-9091]{Michael A. Famiano}
\affiliation{Department of Physics, Western Michigan University, Kalamazoo, Michigan 49008 USA}
\affiliation{Division of Science, National Astronomical Observatory of Japan, 2-21-1 Osawa, Mitaka, Tokyo 181-8588, Japan}

\author[0000-0002-8619-359X]{Toshitaka Kajino}
\affiliation{Division of Science, National Astronomical Observatory of Japan, 2-21-1 Osawa, Mitaka, Tokyo 181-8588, Japan}
\affiliation{School of Physics, Beihang University, 37 Xueyuan Road, Haidian-qu, Beijing
100083, China}
\affiliation{Graduate School of Science, The University of Tokyo, 7-3-1 Hongo, Bunkyo-ku, Tokyo, 113-0033 Japan}
\author[0000-0003-3083-6565]{Motohiko Kusakabe}
\affiliation{School of Physics, Beihang University, 37 Xueyuan Road, Haidian-qu, Beijing 100083, China}
\author[0000-0002-2999-0111]{A. Baha  Balantekin}
\affiliation{Department of Physics, University of Wisconsin-Madison, Madison, Wisconsin 53706 USA}
\affiliation{Division of Science, National Astronomical Observatory of Japan, 2-21-1 Osawa, Mitaka, Tokyo 181-8588, Japan}


\begin{abstract}

Type Ia supernovae (SNe Ia) are believed to be a thermonuclear explosion of a white dwarf, but the mass of their progenitors is still an open problem. In near-Chandrasekhar-mass (near-$M_\mathrm{Ch}$) models of SNe Ia, the central density reaches $\gtrsim10^9\;\mathrm{g\; cm^{-3}}$. The electron chemical potential becomes higher than the $Q$-values of electron capture (EC) transitions between $fp$-shell nuclei, so a portion of the available electrons is captured by iron group elements and thus neutron-rich isotopes are formed. Since EC reaction rates are sensitive to the density, the degree of neutronization is a key to distinguish near- and sub-$M_\mathrm{Ch}$ models. In order to compare observations and theoretical models,  an accurate treatment of EC reactions is necessary. In previous theoretical works, however, effects of electron screening on ECs are ignored. Screening lowers EC rates and thus leads to a higher electron fraction. We implement electron screening on ECs to calculate explosive SN Ia nucleosynthesis in a near-$M_\mathrm{Ch}$ single degenerate model. It is found that some of neutron-rich nuclear abundances, namely those of $^{46,\;48}$Ca, $^{50}$Ti, $^{54}$Cr, $^{58}$Fe, $^{64}$Ni and $^{67,\;70}$Zn, decrease when screening effects on ECs are considered. Of these, $^{50}$Ti, $^{54}$Cr and $^{58}$Fe are particularly interesting because a significant portion of the solar abundance of these nuclei is presumed to originate from SNe Ia. We conclude that implementing the screening effect on ECs in modern SN Ia models is  desirable to precisely calculate abundances of neutron-rich nuclides.

\end{abstract}



\section{Introduction} \label{sec:intro}

Explosive nucleosynthesis in Type Ia supernovae (SNe Ia) is a major source of iron group elements in the Galaxy, but the nature of their progenitor is still under debate. Proposed models of SN Ia progenitors are classified into two regimes. In  near-Chandrasekhar-mass  (near-$M_\mathrm{ch}$) models \cite[e.g.][]{1973ApJ...186.1007W,1984ApJS...54..335I}, carbon fusion is ignited in a white dwarf (WD) when its mass 
gets close to $M_\mathrm{ch}$. On the other hand, in sub-$M_\mathrm{ch}$ models \cite[e.g.][]{1994ApJ...423..371W,2010ApJ...709L..64G}, a SN explosion is triggered even if a WD is lighter than $M_\mathrm{ch}$.

Electron capture (EC) reactions play a key role in
each model. In near $M_\mathrm{ch}$-models, more neutron-rich isotopes are produced because of the high central density $\gtrsim10^9\;\mathrm{g\; cm^{-3}}$. Abundances of neutron-excess isotopes have  been measured in astronomical observations. X-ray observations of SN Ia remnants Kepler \citep{2013ApJ...767L..10P}, Tycho \citep{2014ApJ...780..136Y}, and 3C 397 \citep{2015ApJ...801L..31Y} have been performed to estimate the nickel and manganese abundances. Also, late-time light curves of SN 2011fe \citep{2017ApJ...841...48S,2017MNRAS.468.3798D}, 2012cg \citep{2016ApJ...819...31G}, 2013aa \citep{2018ApJ...857...88J}, 2014J \citep{2018ApJ...852...89Y,2019ApJ...882...30L}, and 2015F \citep{2018ApJ...859...79G} can be used to estimate the abundances of $^{57}$Co and $^{55}$Fe, assuming that they are powered by the decay-chains of 
isobars mass numbers $A=57$ and $A=55$ \citep{2009MNRAS.400..531S}. In order to understand the origin of SN Ia, it is necessary to accurately calculate  abundances of neutron-rich nuclei and compare the models with the observations.

The measurement of EC rates of fully-stripped unstable nuclei is all but impossible, so nuclear shell models have been adopted to calculate them theoretically \citep{1982ApJ...252..715F,1982ApJS...48..279F,1998PhRvC..58..536D,2001ADNDT..79....1L,2004PhRvC..69c4335H,2005JPhCS..20....7H}. Although the shell models have gradually become more sophisticated, only bare EC rates have been used as an input to SN Ia models \cite[e.g.][]{1999ApJS..125..439I,2000ApJ...536..934B,2013A&A...557A...3P,2016ApJ...833..179M,2018ApJ...863..176M,2019A&A...624A.139B}. However, since nuclear reactions in SNe occur in ionized plasma, electron screening changes effective nuclear reaction rates. Such effects have been considered only for thermonuclear charged particle reaction rates \citep{1954AuJPh...7..373S,1973ApJ...181..439D,1973ApJ...181..457G, famiano16}. Effects of screening on ECs have not been estimated in SNe Ia, though they have been explored in big bang nucleosynthesis~\citep{luo20} and SNe II~\citep{famiano20}.  These potentially affect neutronization in SNe Ia remarkably.

We therefore perform calculations of SN Ia nucleosynthesis with screened EC rates. This paper is organized as follows. In Section \ref{method}, we explain the method to calculate the effect of screening on ECs and the adopted SN Ia models. In Section \ref{result}, we compare the results of SN Ia nucleosynthesis calculations with and without electron screening with each other. These results are discussed and summarized in Section \ref{disc}.

\section{Method}
\label{method}
\subsection{Electron Capture Rates}
The EC rates in stellar environments are evaluated as \citep{1982ApJ...252..715F,1982ApJS...48..279F,2001ADNDT..79....1L,2011Suzu}
\begin{eqnarray}
\lambda &=& \frac{\ln 2}{6146 (\mathrm{s})} \sum_{i}W_i \sum_{j}(B_{ij}(GT)+B_{ij}(F))\Phi^{ec}, \nonumber\\
\Phi^{ec} &=& \int_{\omega_\mathrm{min}}^{\infty} \omega p (Q_{ij}+\omega)^2 F(Z, \omega) S_e(\omega) d\omega, \nonumber\\
Q_{ij} &=& (M_p c^2 -M_d c^2 +E_i -E_f )/m_e c^2, \nonumber\\
W_{i} &=& \frac{(2J_i +1) e^{-E_i/kT} }{\sum_{i} (2J_i +1) e^{-E_i/kT}}, 
\end{eqnarray}
where $\omega$($p$) is electron energy (momentum) in units of $m_e c^2$ ($m_e c$), $m_e$ is the electron mass, $M_p$ and $M_d$ are nuclear masses of parent and daughter nuclei, respectively, and $E_i$ ($E_f$) is the excitation energy of initial (final) state.
Here, $B(GT)$ and $B(F)$ are Gamow-Teller and Fermi transition strengths, respectively, given by
\begin{eqnarray}
B_{ij}(GT) &=& \left(\frac{g_A}{g_V}\right)^2 \frac{1}{2J_i +1} |\langle f||\sum_{k} \sigma^k t_{+}^k ||i\rangle|^2, \nonumber\\ 
B_{ij}(F) &=& \frac{1}{2J_i +1} |\langle f||\sum_{k} t_{+}^k ||i\rangle|^2, 
\end{eqnarray} 
where $J_i$ is the total spin of initial state and $t_{+}|p\rangle =|n\rangle$.
$F(Z, \omega)$ is the Fermi function and $S_e(\omega)$ is the Fermi-Dirac distribution for electrons, with the chemical potential determined at high densities, electron fraction, and temperature, indicated by $\rho Y_\mathrm{e}$,  $Y_\mathrm{e}$, and $T$, respectively.  The chemical potential is determined by,
\begin{eqnarray}
\rho Y_\mathrm{e} &=& \frac{1}{\pi^2 N_A}\left(\frac{m_e c}{\hbar}\right)^3 \int_{0}^{\infty} (S_e -S_p) p^2 dp, \nonumber\\
S_{\ell} &=& \frac{1}{\exp(\frac{E_\ell -\mu_\ell}{kT})+1},
\end{eqnarray}
where  $\ell=e$ for electrons and $p$ for positrons and $\mu_p = -\mu_e$ is the positron chemical potential. 

Here, the Coulomb corrections on the transition rates due to the electron background are studied. 
The screening effects on both electrons and ions are taken into account for the Coulomb effects \citep{2010Juod,2013Toki,2016Suzu}. 
The screening effects of electrons are evaluated by using the dielectric function obtained by relativistic random phase approximation (RPA) \citep{2002Itoh}.
The effect is included by reducing the chemical potential of electrons by an amount equal to the modification of the Coulomb potential at the origin $V_s(0)$ \citep{2010Juod}, where
\begin{eqnarray}
V_s(r) &=& Ze^2 (2k_F) J(r), \nonumber\\
J(r) &=& \frac{1}{2k_F r}\left(1-\frac{2}{\pi}\int \frac{\sin(2k_F qr)}{q \epsilon (q,0)} dq\right). 
\end{eqnarray}
Here, $\epsilon (q,0)$ is the static longitudinal dielectric function at zero frequency, and $q=k/2k_F$ with $k$ and $k_F$ the electron wave number and Fermi wave number, respectively.
The modification to the Coulomb potential $J(r)$ is tabulated in \citet{2002Itoh}.

The other Coulomb effect is caused by the screening of the ions in the electron background.
The threshold energy is modified by
\begin{equation}
\Delta Q_C = \mu_C (\mbox{Z-1}) -\mu_C (\mbox{Z}),
\end{equation}
where $\mu_C$ (Z) is the Coulomb chemical potential of the nucleus with charge number Z \citep{1982Slattery,1993Ichimaru}.
The Coulomb chemical potential in a plasma of electron number density $n_e$ and temperature $T$ is given by
\begin{equation}
\mu_C (\mbox{Z}) = kT f(\Gamma),
\end{equation}     
with $\Gamma$ = Z$^{5/3}$ $\Gamma_e$, $\Gamma_e = \frac{e^2}{kT a_e}$ and $a_e$ = $(\frac{3}{4\pi n_e})^{1/3}$. 
The function $f$ for the strong-coupling regime, $\Gamma >1$, is given by Equation (A.48) in \citet{1993Ichimaru},
while for the weak-coupling regime, $\Gamma <1$,  an analytic function given by \citet{1989ASPRv...7..311Y} is used  \citep[see also Equation (A.6) in][]{2010Juod}.
The threshold energy is enhanced for EC processes, and the EC ($\beta$-decay) rates are reduced (enhanced) by the Coulomb effects. 
  
\begin{figure}
\centering
\includegraphics[width=8.5cm]{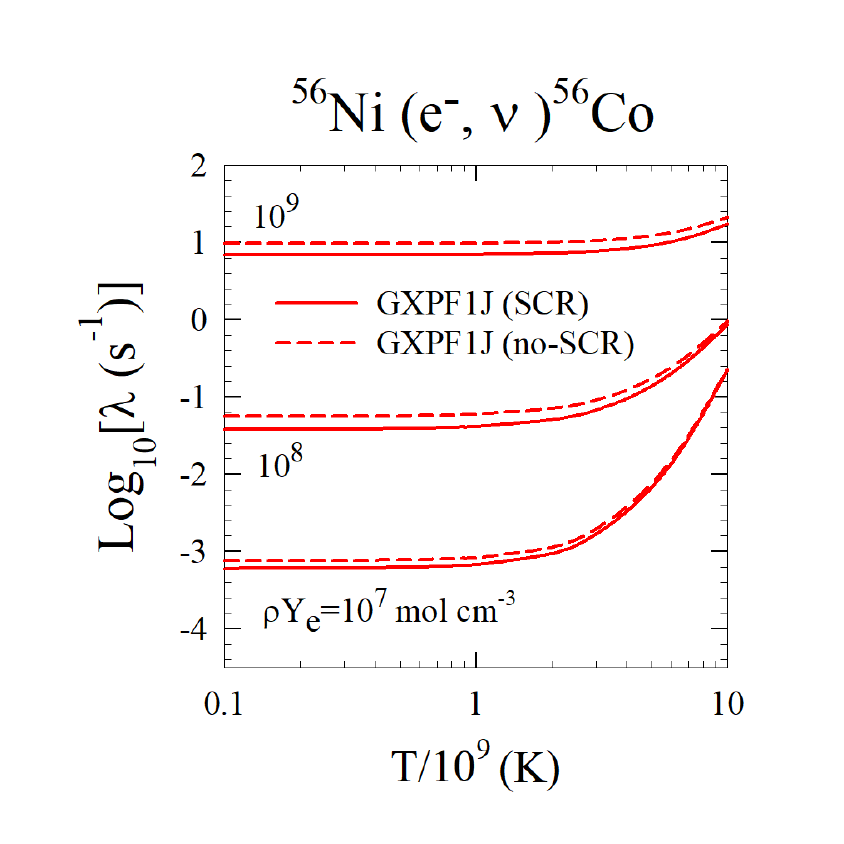}
\caption{Comparison of calculated e-capture rates for $^{56}$Ni (e$^{-}$, $\nu$) $^{56}$Co  obtained with the GXPF1J at densities $\rho Y_\mathrm{e} = 10^{7}$, 10$^{8}$ and 10$^{9}$ mol cm$^{-3}$ for temperatures $T = 10^{8}$-$10^{10}$ K. 
Solid and dashed curves denote the rates with and without the screening effects, respectively.}
\label{fig:rate56}
\end{figure}

The EC rates for $pf$-shell in stellar environments are evaluated with the use of the shell-model Hamiltonian, GXPF1J \citep{2005JPhCS..20....7H}, which is a modified version of GXPF1 \citep{2004PhRvC..69c4335H}.  
The quenching of the axial-vector coupling constant is taken to be $g_A^{eff}/g_A$ = 0.74 \citep{2005Caurier}. 
Transitions from the states with excitation energies up to $E_X = 2$ MeV are taken into account.
Here, the experimental data such as excitation energies for excited states in both parent and daughter nuclei and $B(GT)$ values are taken into account when they are available in the online retrival system of National Nuclear Data Center\footnote{https://www.nndc.bnl.gov/}.
Calculated e-capture rates for $^{56}$Ni (e$^{-}$, $\nu$) $^{56}$Co with and without the screening effects are shown in Fig. \ref{fig:rate56}. 
The weak rates for the case with the screening effects are found to be reduced by about 20-40$\%$ compared with those without the screening effects.  

\subsection{Nucleosynthesis Calculation}
\begin{figure}
\centering
\includegraphics[width=8.5cm]{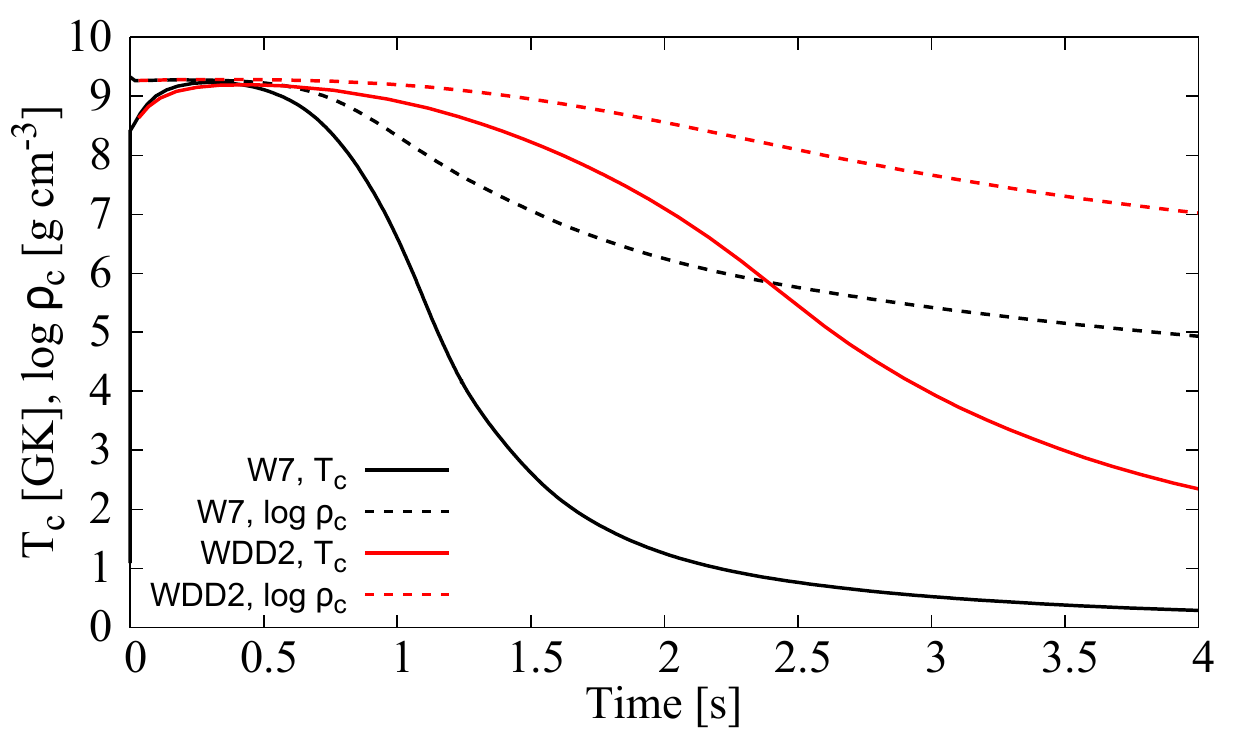}
\caption{The time evolution of the central temperature $T_\mathrm{c}$ and density $\rho_\mathrm{c}$. The black lines show the W7 model and the red lines show the WDD2 model. The solid lines show $T_\mathrm{c}$ and the broken lines show $\rho_\mathrm{c}$.}
\label{fig:traj}
\end{figure}

Because nucleosynthesis in sub-$M_\mathrm{ch}$ models is nearly as sensitive to EC rates \citep{2019A&A...624A.139B}, we focus on near-$M_\mathrm{ch}$ models. Physical mechanism of propagation of the burning front in SNe Ia is still unclear \citep[e.g.][]{1999ApJ...523L..57N,2017suex.book.....B}. We adopt two one-dimensional near-$M_\mathrm{ch}$ models with different burning schemes. One is W7 \citep{1984ApJ...286..644N}, which is a widely-used deflagration model, and the other is WDD2 \citep{1999ApJS..125..439I}, which is a delayed-detonation model. These models adopt the equation of state in \citet{1982Slattery} and  \citet{1980PhRvA..21.2087S}, which take the Coulomb correction into account. Since the critical density of the deflagration-detonation transition is not known \citep{1997NewA....2..239N}, it is assumed to be $2.2\times10^7\;\mathrm{g\;cm^{-3}}$ in WDD2. The time evolution of the central temperature $T_\mathrm{c}$ and density $\rho_\mathrm{c}$ of the models is shown in Fig. \ref{fig:traj}.

We performed nuclear reaction network calculations as a post process with W7 and WDD2 using  \texttt{libnucnet} \citep{2007M&PSA..42.5215M}. The initial composition is $X(^{12}\mathrm{C})=0.475$, $X(^{16}\mathrm{O})=0.50$, and $X(^{22}\mathrm{Ne})=0.025$, where $X(i)$ is the mass fraction of each nucleus $i$. The network calculation is performed until 100 s after the ignition at the WD center and after that unstable nuclei are forced to decay. The network includes 5441 nuclear species up to astatine. Thermonuclear reaction rates are taken from the JINA REACLIB v1.1 database \citep{2010ApJS..189..240C}. EC rates of $fp$-shell nuclei are calculated from the GXPF1J shell model \citep{2005JPhCS..20....7H}.  The EC rates of other nuclei are taken from \citet{1994ADNDT..56..231O} and \citet{1982ApJ...252..715F,1982ApJS...48..279F}. The treatment of electron screening for thermonuclear reactions is based on \citet{1982ApJ...258..696W}. Screened EC rates for $pf$-shell nuclei are treated as we explain in Section 2.1, and  the screening effect on EC rates for other nuclei is not considered. The EC rates are tabulated in a range $\rho Y_e = 10^5-10^{11}$ mol cm$^{-3}$ and $T=10^7-10^{11}$ K \citep{2020HS}. The \texttt{libnucnet} calculates the effective $ft$-values \citep{1985ApJ...293....1F} from the tabulated rates and interpolate $\log ft$ as a linear function of $T$ and $\log\rho$ to perform the network calculation.
\section{Results}
\label{result}
Using the modifications to the EC capture rates previously
computed, the W7 and WDD2 thermodynamic trajectories were used 
to compute the final nuclear abundances in each model.
\subsection{W7 Model}
\label{w7}
\begin{figure}
\centering
\includegraphics[width=8.5cm]{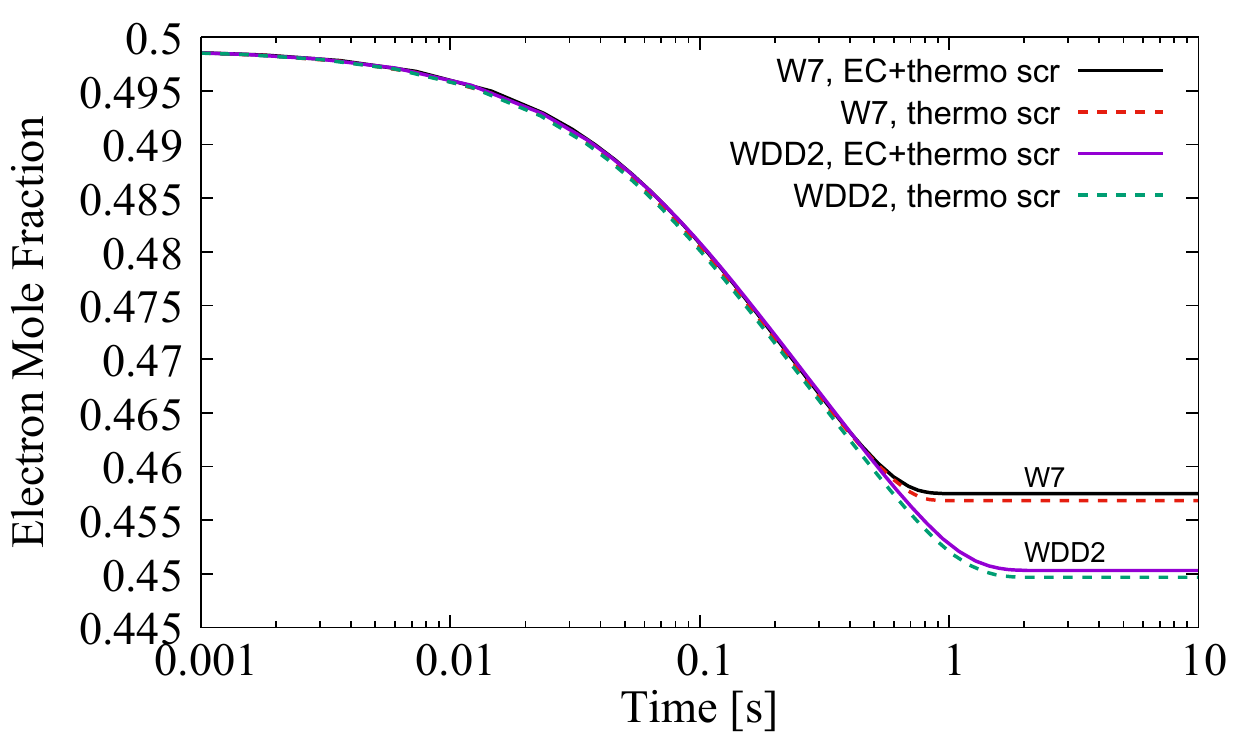}
\caption{The time evolution of the electron fraction  at the center of the SN Ia models. The solid lines indicate the result with the screening effect on both of ECs and thermonuclear reactions, while the broken lines indicate the result with the screening effect only on thermonuclear reactions. }
\label{fig:ye}
\end{figure}
The neutronization degree of plasma can be represented by the electron fraction $Y_\mathrm{e}=\sum_iZ_iY_i$, where $Z_i$ is the atomic number and $Y_i$ is the mole fraction of the $i$-th nuclear species. Thus, for equal numbers of neutrons and protons, the $Y_\mathrm{e}$ is exactly 0.5. If the number of neutrons exceeds that of protons, the $Y_\mathrm{e}$ becomes lower than 0.5.

Fig. \ref{fig:ye} shows the time evolution of $Y_\mathrm{e}$ at the center of the explosion. The solid lines indicate the results with the screening effect on both of ECs and thermonuclear reactions, while the broken lines indicate the results with the screening effect only on thermonuclear reactions. 

In our models, the initial electron fraction is $Y_\mathrm{e}=0.49886$, which 
is slightly lower than 0.5 because of the abundance of the neutron-rich nucleus
$^{22}$Ne in the initial composition. The EC reactions freeze out after $\sim1$
s of the explosion. It is seen that the screened lower EC rates results in a 
higher $Y_\mathrm{e}$. The electron fraction at 10 s is $Y_\mathrm{e}=0.45747$ 
if screening on ECs is considered, while $Y_\mathrm{e}=0.45682$ if screening only 
on thermonuclear reactions is considered.
\begin{figure}
\centering
\includegraphics[width=8.5cm]{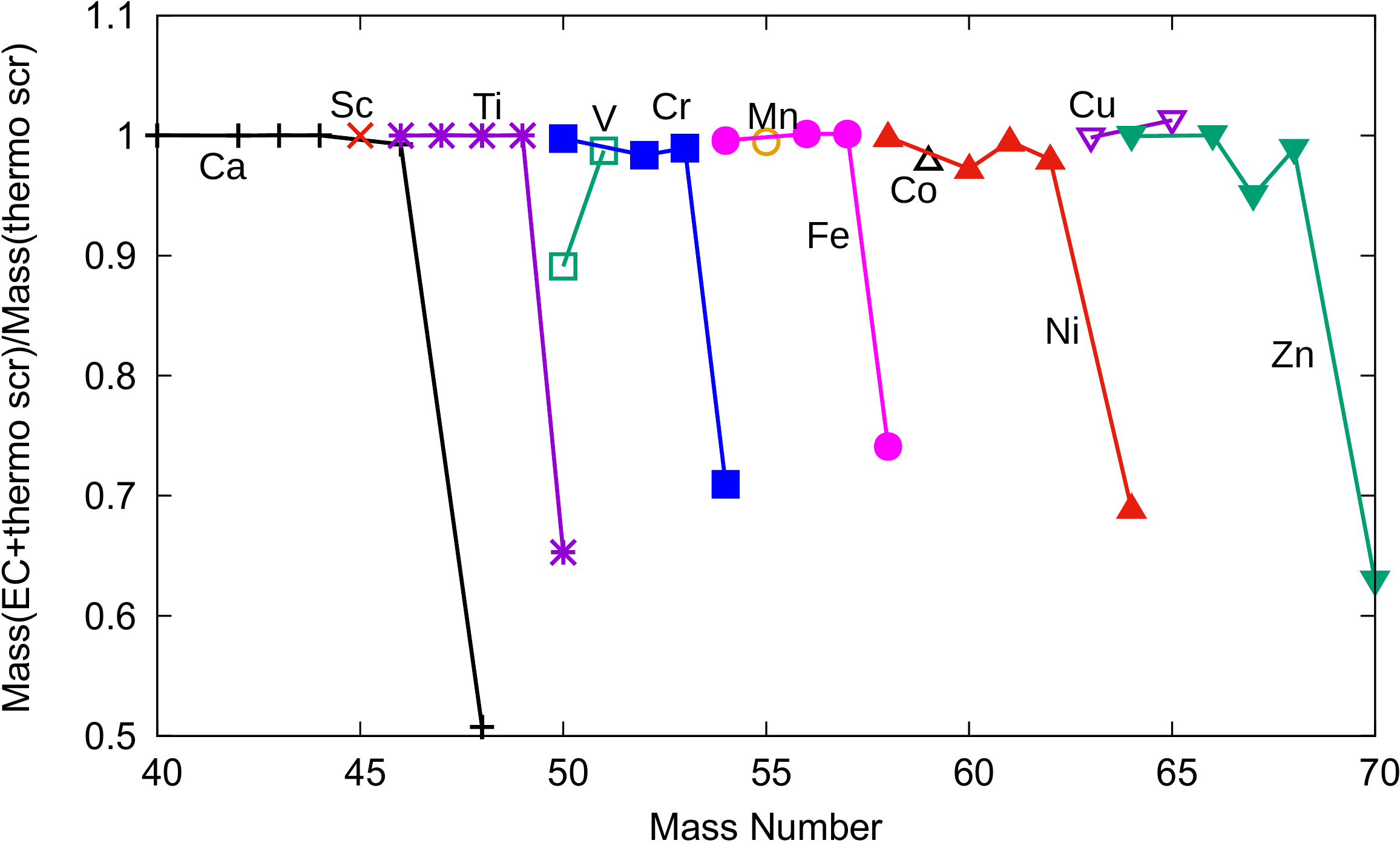}
\caption{The abundance ratios with and without screening on ECs. W7 is adopted as a SN Ia model.}
\label{fig:ratio_w7}
\end{figure}

It is known that neutronization in the central region of a WD is determined by the EC on protons \citep[e.g.][]{2019A&A...624A.139B}. Although the screening effect on the EC of protons is not considered in the present work, we checked that the change in the EC rate of protons is within $1\%$. This is because the electric charge of protons is smaller than those of $fp$-shell nuclei. When we consider the screening effect on free protons, the central electron fraction is $Y_\mathrm{e}=0.45751$. This value is slightly higher than the one shown in the previous paragraph, but it does not lead to qualitative differences in abundances.

Fig. \ref{fig:ratio_w7} shows a comparison of the nucleosynthetic yields in the
cases with and without the screening effect on ECs. One can find that the 
abundances of neutron-rich isotopes are smaller when the screening effect on 
ECs is considered. Most notably, the abundances of $^{48}$Ca, $^{50}$Ti, 
$^{54}$Cr, $^{58}$Fe, $^{64}$Ni and $^{70}$Zn are 30-50\% smaller.
\begin{figure}
\centering
\includegraphics[width=8.5cm]{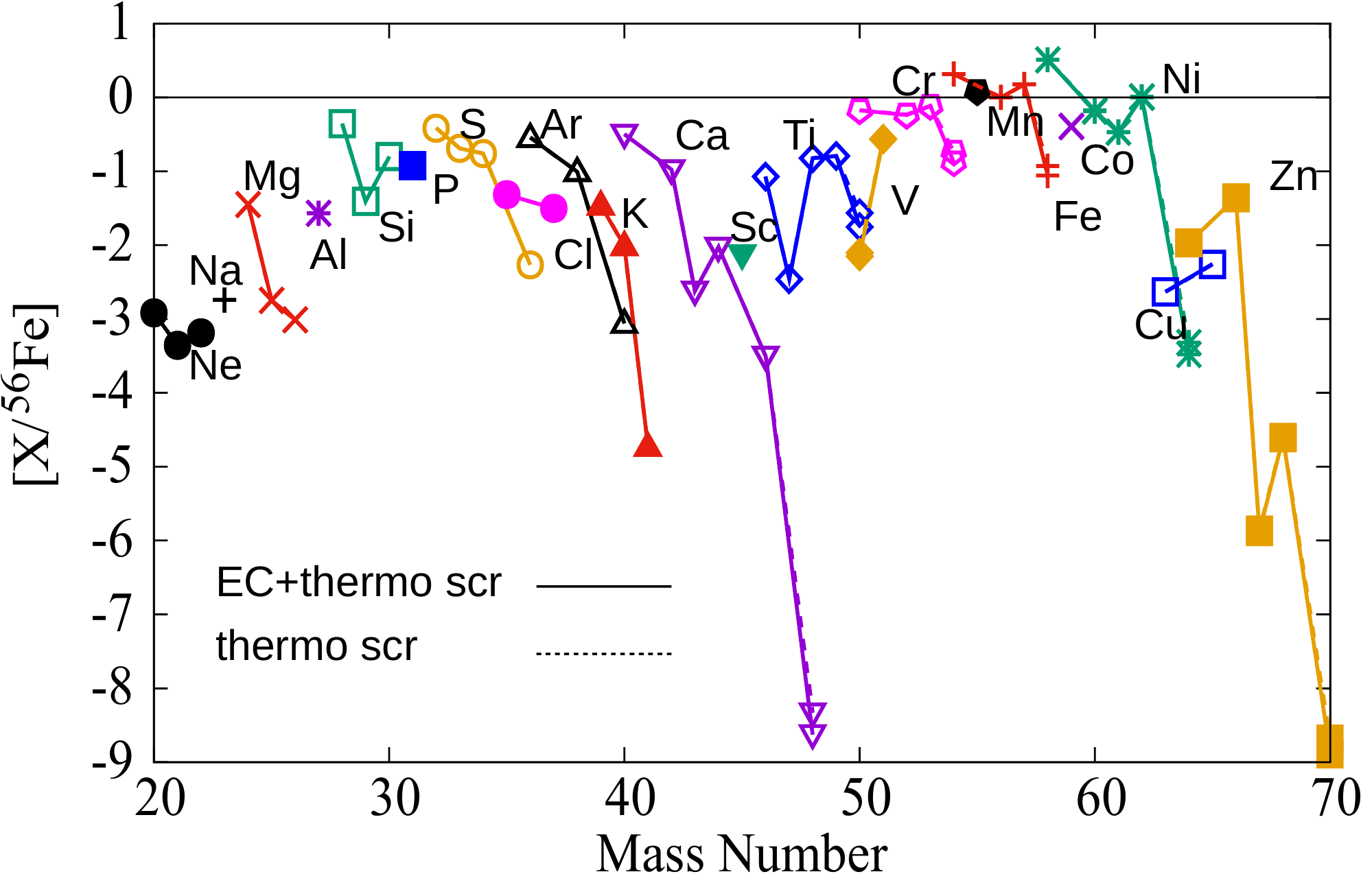}
\caption{The abundances normalised by the Solar and $^{56}$Fe abundances. The solid lines adopt screening on ECs and the broken lines do not. W7 is adopted as a SN Ia model.}
\label{fig:sol_w7}
\end{figure}

Fig. \ref{fig:sol_w7} shows the SN Ia yields normalized to the solar abundances. The normalized abundance  $[X/^{56}\mathrm{Fe}]$  for a nuclide $X$ is defined as
\begin{equation}
[X/^{56}\mathrm{Fe}]=\log\left(\frac{Y(X)}{Y(^{56}\mathrm{Fe})}\right)-\log\left(\frac{Y_\odot(X)}{Y_\odot(^{56}\mathrm{Fe})}\right),
\end{equation}
where $Y(i)$ and $Y_\odot(i)$ are respectively the number abundances in the SNe Ia models and  in the Solar System of species $i$ \citep{2019arXiv191200844L}. The solid lines show 
the result in the case with the screening effect on  ECs, while the dashed 
lines indicate results without EC screening. In a SN Ia, as much as 
$\sim0.8M_\odot$ of iron group elements can be produced, while only 
$\sim0.1M_\odot$ is produced in a core-collapse SN. For a Galactic SNe Ia event 
rate of ($\sim0.54\pm0.12$) /century and a core-collapse SNe event rate of ($\sim2.30\pm0.48$)
/century \citep{2011MNRAS.412.1473L}, it is expected that $\sim65\%$ of the 
iron group elements originates from SNe Ia. For the neutron-rich nuclides, 
production is affected by screening of ECs. The contribution of SNe Ia
to the Solar abundance of $^{48}$Ca, $^{50}$Ti, $^{64}$Ni and $^{70}$Zn is
negligible. However, $[^{54}\mathrm{Cr}/^{56}\mathrm{Fe}]$ and
$[^{58}\mathrm{Fe}/^{56}\mathrm{Fe}]$ reach $\sim-1$. This implies that the
contribution of SNe Ia for these nuclei can be as high as $\sim$5-10\%. The
contribution of SNe Ia to the Solar abundances of these nuclei is summarized in
Table \ref{w7_tab}. The abundances of $^{54}$Cr and $^{58}$Fe are significantly
affected by screening of ECs. It is therefore remarkably
important to consider its effect in discussing the origins of these nuclei.
\begin{table}[t]
\centering
 \begin{tabular}{cccc}
 &$^{50}$Ti&$^{54}$Cr&$^{58}$Fe\\\hline\hline
 EC+thermo scr&$1.2\pm0.2$&$8.5\pm1.2$&$5.7\pm0.8$\\\hline
thermo scr&$1.8\pm0.3$&$12\pm2$&$7.6\pm1.1$\\\hline
\end{tabular}
\caption{Contribution of SNe Ia to the Solar abundances in units of percent. The W7-like explosion is assumed to dominate the whole SN Ia population. The uncertainties are estimated based on errors in Galactic SN rates \citep{2011MNRAS.412.1473L}.}
\label{w7_tab}
\end{table}

\subsection{WDD2 Model}
\label{wdd2}
The qualitative effect of screening  on the WDD2 model is similar to that of the W7 model. Fig. \ref{fig:ye} shows the time evolution of $Y_\mathrm{e}$. As seen in W7, the resultant $Y_\mathrm{e}$ is higher as a result of EC screening. The electron fraction at $t=10$ s is $Y_\mathrm{e}=0.45031$ if screening effects on ECs and thermonuclear reactions are considered and $Y_\mathrm{e}=0.44967$ if screening only on thermonuclear reactions is considered. 
\begin{figure}
\centering
\includegraphics[width=8.5cm]{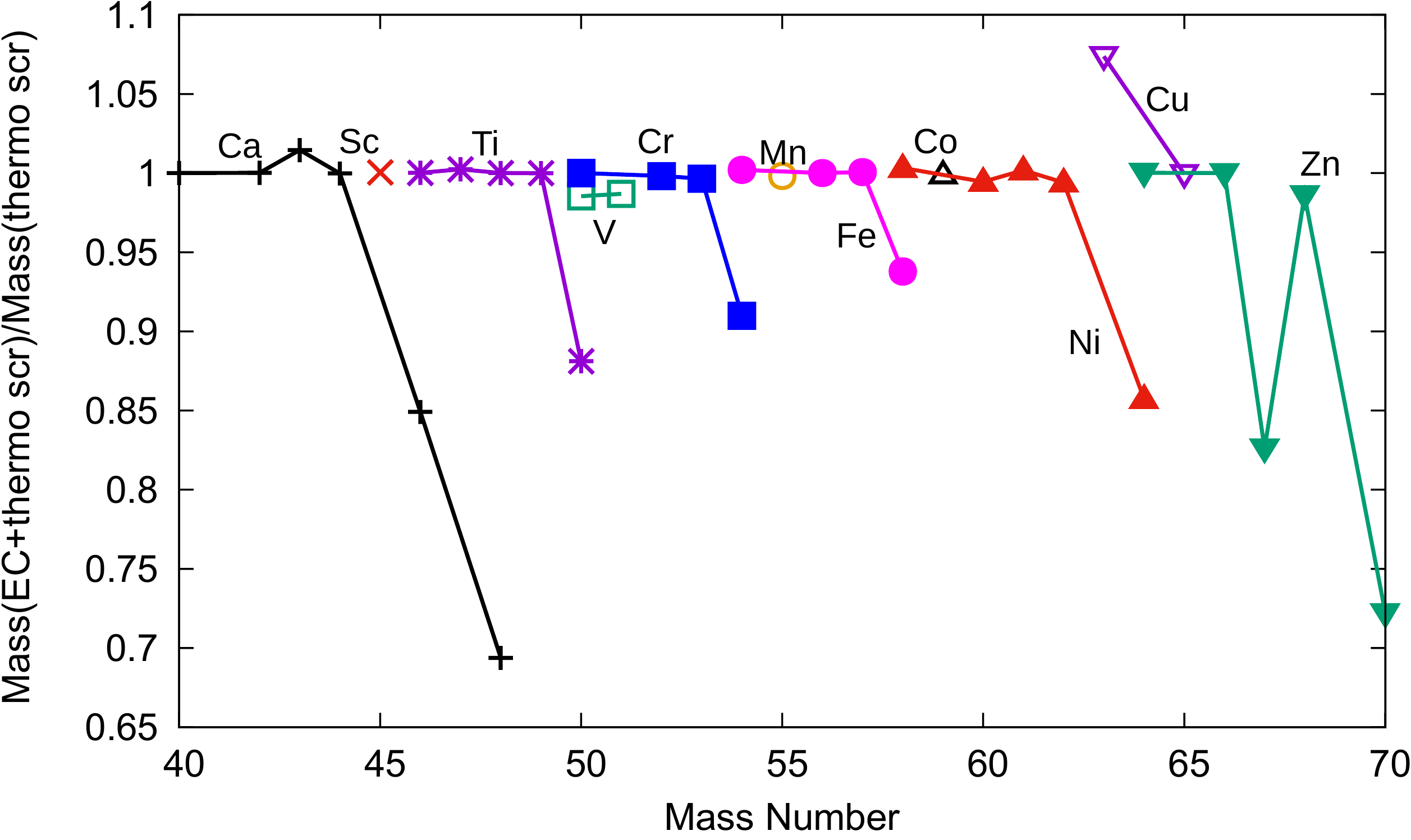}
\caption{The abundance ratios between the cases with and without screening on ECs. WDD2 is adopted as a SN Ia model.}
\label{fig:ratio_wdd2}
\end{figure}

Fig. \ref{fig:ratio_wdd2} shows the abundance ratio for cases with and without screening on ECs. It is seen that the abundances of neutron-rich isotopes tend to be smaller if screening on ECs is considered. In particular, the abundances of $^{46}$Ca, $^{48}$Ca, $^{50}$Ti, $^{54}$Cr, $^{58}$Fe, $^{64}$Ni, $^{67}$Zn and $^{70}$Zn are 10-30\% smaller.
\begin{figure}
\centering
\includegraphics[width=8.5cm]{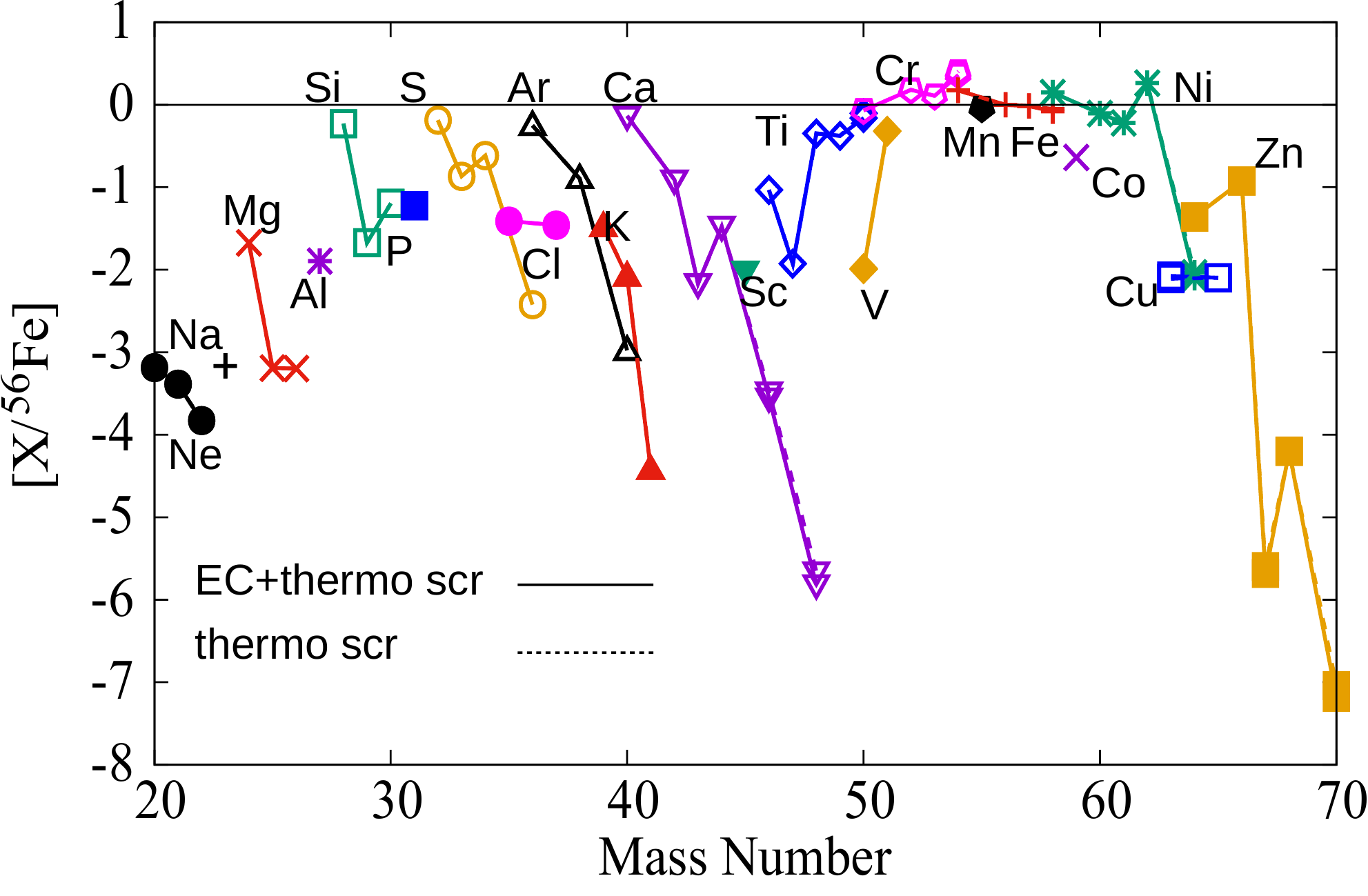}
\caption{The abundances normalised by the Solar and $^{56}$Fe abundances. The solid lines adopt screening on ECs and the broken lines do not. WDD2 is adopted as a SN Ia model.}
\label{fig:sol_wdd2}
\end{figure}

While the final abundances of almost all nuclei are not affected by the suppressed EC rates, the abundance of $^{63}$Cu increases by 7\% when screening on ECs is considered. Half of $^{63}$Cu is formed by the $\beta$-decay of $^{63}$Zn whose life-time is 38 min. Because excess of proton number of $^{63}$Zn is larger than any stable isotopes of zinc, its abundance increases by the suppressed EC rates. As a result, the abundance of $^{63}$Cu becomes higher.
\begin{table}[t]
\centering
 \begin{tabular}{cccc}
 &$^{50}$Ti&$^{54}$Cr&$^{58}$Fe\\\hline\hline
 EC+thermo scr&$45\pm7$&$150\pm22$&$54\pm8$\\\hline
thermo scr&$51\pm8$&$165\pm24$&$58\pm9$\\\hline
\end{tabular}
\caption{Contribution of SNe Ia to the Solar abundances in units of percent. The WDD2-like explosion is assumed to dominate the whole SN Ia population. The uncertainties are estimated based on errors in Galactic SN rates \citep{2011MNRAS.412.1473L}.}
\label{wdd2_tab}
\end{table}

Fig. \ref{fig:sol_wdd2} shows the abundances normalized to Solar values. The solid lines indicate results corresponding to the case with EC screening, and the dashed lines correspond to screening of thermonuclear reactions only. For the nuclides affected by EC screening, the contribution of SNe Ia to the Solar abundances for $^{46,\;48}$Ca, $^{64}$Ni, and $^{67,\;70}$Zn is negligibly small. However, the abundances of $^{50}$Ti, $^{54}$Cr, and $^{58}$Fe reach $[^{50}\mathrm{Ti}/^{56}\mathrm{Fe}]=-0.16$, $[^{54}\mathrm{Cr}/^{56}\mathrm{Fe}]=0.36$, and $[^{58}\mathrm{Fe}/^{56}\mathrm{Fe}]=-0.080$, where screening on ECs is considered. The SN Ia contribution to the Solar abundances can be as high as $\sim50\%$ for $^{50}$Ti and $^{58}$Fe, and $>100\%$ for $^{54}$Cr. The overproduction of $^{54}$Cr has been reported by previous works \citep{1999ApJS..125..439I,2016ApJ...833..179M,2018ApJ...861..143L} as well. The contribution of SNe Ia for these nuclei is summarized in Table \ref{wdd2_tab}.

\section{Discussion}
\label{disc}
In this study, we calculate EC rates of $pf$-shell nuclei and apply them to SN Ia nucleosynthesis. The suppressed EC rates result in smaller $Y_\mathrm{e}$ values in SN ejecta and thus smaller abundances of  neutron-rich nuclei. The abundances of $^{46,\;48}$Ca, $^{50}$Ti, $^{54}$Cr, $^{58}$Fe, $^{64}$Ni and $^{67,\;70}$Zn are most prominently affected, although the details of the abundance pattern and sensitivity to the EC rates depend on the explosion models. This result is consistent with previous works \citep{2000ApJ...536..934B,2019A&A...624A.139B} which point out strong sensitivity of these nuclei to EC rates. 

Screening of ECs does not affect the abundances of isobars with mass numbers of $A=55$ and 57, which are estimated from late-time light curves of SNe Ia. The effect on the elemental abundances of nickel and manganese, which are estimated in SN remnants, are $\lesssim 0.4\%$ for nickel and $\lesssim0.5\%$ for manganese.

Of the neutron-rich nuclei that are affected by screening on ECs, $^{50}$Ti, $^{54}$Cr and $^{58}$Fe are particularly interesting because SNe Ia can significantly contribute to their Solar abundances. Since the production of these nuclei depends on the central density \citep{2013MNRAS.429.1156S,2018ApJ...861..143L}, they are a good indicator of the mass of SN Ia progenitors. Hence the information of SN Ia models can be imprinted in the Solar abundance patterns of titanium, chromium, and iron. In order to compare the SN Ia models and the observed abundance patterns of these nuclei, we recommend implementing the screened EC rates in modern multi-dimensional SN Ia models \citep{2010ApJ...712..624M,2010ApJ...720...99J,2012ApJ...757..175K,2013MNRAS.429.1156S,2018ApJ...861..143L}. 

The post-processing technique adopted in this study decouples hydrodynamics from detailed nucleosynthesis. However, since the electron pressure depends on $Y_\mathrm{e}$, changes in EC rates can significantly affect overall dynamics of SN Ia. For example, incinerated bubbles around the WD center float outwards in three-dimensional models, and their motion is dependent on EC rates \citep{2019A&A...624A.139B}. Also, the spatial extent of the density inversion behind the deflagration front is dependent on  EC rates \citep{1992ApJ...396..649T}. These interplays between dynamics and nuclear reactions are not taken into account in the current post-process calculation for one-dimensional explosion models. It is hence desirable to couple hydrodynamics and the updated EC rates and consider multi-dimensionality in future studies. 

\acknowledgments

This work was supported by JSPS KAKENHI Grant Number JP19J12892 and JP19K03855. T.K. is supported in part by Grands-in-Aid for Scientific Research of JSPS Grant No. 20K03958 and 17K05459.   
A.B.B. is supported in part by the U.S. National Science Foundation Grant No. PHY-1806368. M.A.F. is supported by NASA grant \#80NSSC20K0498. M.A.F. and A.B.B. acknowledge support from the NAOJ Visiting Professor program.
The authors would like to thank IReNA project for the promotion of the compilation and update of nuclear weak rates.
%

\vspace{5mm}


\software{Libnucnet \citep{2007M&PSA..42.5215M}  
          }







\begin{thebibliography}{dummy} 
  \bibitem[Brachwitz et al.(2000)]{2000ApJ...536..934B} Brachwitz, F., Dean, D.~J., Hix, W.~R., et al.\ 2000, \apj, 536, 934
  \bibitem[Branch \& Wheeler(2017)]{2017suex.book.....B} Branch, D., \& Wheeler, J.~C.\ 2017, Supernova Explosions: Astronomy and Astrophysics Library

\bibitem[Bravo(2019)]{2019A&A...624A.139B} Bravo, E.\ 2019, \aap, 624, A139
\bibitem[Caurier et al.(2005)]{2005Caurier} Caurier, E., Mart{\'\i}nez-Pinedo, G., Nowacki, F., Poves, A., \& Zuker, A.~P. \ 2005, \rmp, 77, 427
\bibitem[Cyburt et al.(2010)]{2010ApJS..189..240C} Cyburt, R.~H., Amthor, A.~M., Ferguson, R., et al.\ 2010, \apjs, 189, 240

  \bibitem[Dean et al.(1998)]{1998PhRvC..58..536D} Dean, D.~J., Langanke, K., Chatterjee, L., et al.\ 1998, \prc, 58, 536
  \bibitem[Dewitt, Graboske \& Cooper(1973)]{1973ApJ...181..439D} Dewitt, H.~E., Graboske, H.~C., \& Cooper, M.~S.\ 1973, \apj, 181, 439

\bibitem[Dimitriadis et al.(2017)]{2017MNRAS.468.3798D} Dimitriadis, G., Sullivan, M., Kerzendorf, W., et al.\ 2017, \mnras, 468, 3798
\bibitem[Famiano, Balantekin, \& Kajino(2016)]{famiano16}Famiano, M.A., Balantekin, A.B., \& Kajino, T.\ 2016 \prc, 93, 045804
\bibitem[Famiano et al.(2020)]{famiano20} Famiano, M., Balantekin, A.~B., Kajino, T., et al.\ 2020, \apj, 898, 163

\bibitem[Fuller, Fowler \& Newman(1982a)]{1982ApJ...252..715F} Fuller, G.~M., Fowler, W.~A., \& Newman, M.~J.\ 1982, \apj, 252, 715
\bibitem[Fuller, Fowler \& Newman(1982b)]{1982ApJS...48..279F} Fuller, G.~M., Fowler, W.~A., \& Newman, M.~J.\ 1982, \apjs, 48, 279
\bibitem[Fuller, Fowler \& Newman(1985)]{1985ApJ...293....1F} Fuller, G.~M., Fowler, W.~A., \& Newman, M.~J.\ 1985, \apj, 293, 1

\bibitem[Graboske et al.(1973)]{1973ApJ...181..457G} Graboske, H.~C., Dewitt, H.~E., Grossman, A.~S., et al.\ 1973, \apj, 181, 457

  \bibitem[Graur et al.(2016)]{2016ApJ...819...31G} Graur, O., Zurek, D., Shara, M.~M., et al.\ 2016, \apj, 819, 31
  \bibitem[Graur et al.(2018)]{2018ApJ...859...79G} Graur, O., Zurek, D.~R., Rest, A., et al.\ 2018, \apj, 859, 79

  \bibitem[Guillochon et al.(2010)]{2010ApJ...709L..64G} Guillochon, J., Dan, M., Ramirez-Ruiz, E., et al.\ 2010, \apjl, 709, L64
  \bibitem[Honma et al.(2004)]{2004PhRvC..69c4335H} Honma, M., Otsuka, T., Brown, B.~A., et al.\ 2004, \prc, 69, 034335

\bibitem[Honma et al.(2005)]{2005JPhCS..20....7H} Honma, M., Otsuka, T., Mizusaki, T., et al.\ 2005, Journal of Physics Conference Series, 7
\bibitem[Honma \& Suzuki(2020)]{2020HS} Honma, M. \& Suzuki, T. \ 2020, https://www.phys.chs.nihon-u.ac.jp/suzuki/data5/, and supplementary data material

  \bibitem[Iben \& Tutukov(1984)]{1984ApJS...54..335I} Iben, I., \& Tutukov, A.~V.\ 1984, \apjs, 54, 335
\bibitem[Ichimaru(1993)]{1993Ichimaru} Ichimaru, S. \ 1993, \rmp, 65, 255
\bibitem[Itoh et al.(2002)]{2002Itoh} Itoh, N., Tomizawa, N., Tamamura, M., \& Wanajo, S. \ 2002, \apj, 579, 380

\bibitem[Iwamoto et al.(1999)]{1999ApJS..125..439I} Iwamoto, K., Brachwitz, F., Nomoto, K., et al.\ 1999, \apjs, 125, 439
\bibitem[Jackson et al.(2010)]{2010ApJ...720...99J} Jackson, A.~P., Calder, A.~C., Townsley, D.~M., et al.\ 2010, \apj, 720, 99

  \bibitem[Jacobson-Gal{\'a}n et al.(2018)]{2018ApJ...857...88J} Jacobson-Gal{\'a}n, W.~V., Dimitriadis, G., Foley, R.~J., et al.\ 2018, \apj, 857, 88
 \bibitem[Juodagalvis et al.(2010)]{2010Juod} Juodagalvis, A., Langanke, K., Hix, W.~R., Mart{\'\i}nez-Pinedo, G., \& Sampaio, J.~M. \ 2010, Nucl. Phys. A, 848, 454

  \bibitem[Krueger et al.(2012)]{2012ApJ...757..175K} Krueger, B.~K., Jackson, A.~P., Calder, A.~C., et al.\ 2012, \apj, 757, 175

    \bibitem[Langanke \& Mart{\'\i}nez-Pinedo(2001)]{2001ADNDT..79....1L} Langanke, K., \& Mart{\'\i}nez-Pinedo, G.\ 2001, Atomic Data and Nuclear Data Tables, 79, 1
    \bibitem[Leung \& Nomoto(2018)]{2018ApJ...861..143L} Leung, S.-C., \& Nomoto, K.\ 2018, \apj, 861, 143

\bibitem[Li et al.(2011)]{2011MNRAS.412.1473L} Li, W., Chornock, R., Leaman, J., et al.\ 2011, \mnras, 412, 1473

\bibitem[Li et al.(2019)]{2019ApJ...882...30L} Li, W., Wang, X., Hu, M., et al.\ 2019, \apj, 882, 30
\bibitem[Lodders(2020)]{2019arXiv191200844L}  Lodders, K. 2020, Solar Elemental Abundances, in The Oxford Research Encyclopedia of Planetary Science, Oxford University Press
\bibitem[Luo et al.(2020)]{luo20}Luo et al.\ 2020, \prd, 101, 083010
\bibitem[Maeda et al.(2010)]{2010ApJ...712..624M} Maeda, K., R{\"o}pke, F.~K., Fink, M., et al.\ 2010, \apj, 712, 624

\bibitem[Meyer \& Adams(2007)]{2007M&PSA..42.5215M} Meyer, B.~S., \& Adams, D.~C.\ 2007, Meteoritics and Planetary Science Supplement, 42, 5215
\bibitem[Mori et al.(2016)]{2016ApJ...833..179M} Mori, K., Famiano, M.~A., Kajino, T., et al.\ 2016, \apj, 833, 179
\bibitem[Mori et al.(2018)]{2018ApJ...863..176M} Mori, K., Famiano, M.~A., Kajino, T., et al.\ 2018, \apj, 863, 176
\bibitem[Niemeyer \& Kerstein(1997)]{1997NewA....2..239N} Niemeyer, J.~C., \& Kerstein, A.~R.\ 1997, \na, 2, 239
\bibitem[Niemeyer(1999)]{1999ApJ...523L..57N} Niemeyer, J.~C.\ 1999, \apjl, 523, L57

\bibitem[Nomoto, Thielemann \& Yokoi(1984)]{1984ApJ...286..644N} Nomoto, K., Thielemann, F.-K., \& Yokoi, K.\ 1984, \apj, 286, 644
\bibitem[Oda et al.(1994)]{1994ADNDT..56..231O} Oda, T., Hino, M., Muto, K., et al.\ 1994, Atomic Data and Nuclear Data Tables, 56, 231
\bibitem[Parikh et al.(2013)]{2013A&A...557A...3P} Parikh, A., Jos{\'e}, J., Seitenzahl, I.~R., et al.\ 2013, \aap, 557, A3
\bibitem[Park et al.(2013)]{2013ApJ...767L..10P} Park, S., Badenes, C., Mori, K., et al.\ 2013, \apjl, 767, L10

\bibitem[Salpeter(1954)]{1954AuJPh...7..373S} Salpeter, E.~E.\ 1954, Australian Journal of Physics, 7, 373
\bibitem[Seitenzahl, Taubenberger, \& Sim(2009)]{2009MNRAS.400..531S} Seitenzahl, I.~R., Taubenberger, S., \& Sim, S.~A.\ 2009, \mnras, 400, 531

\bibitem[Seitenzahl et al.(2013)]{2013MNRAS.429.1156S} Seitenzahl, I.~R., Ciaraldi-Schoolmann, F., R{\"o}pke, F.~K., et al.\ 2013, \mnras, 429, 1156

\bibitem[Shappee et al.(2017)]{2017ApJ...841...48S} Shappee, B.~J., Stanek, K.~Z., Kochanek, C.~S., et al.\ 2017, \apj, 841, 48
\bibitem[Slattery et al.(1980)]{1980PhRvA..21.2087S} Slattery, W.~L., Doolen, G.~D., \& Dewitt, H.~E.\ 1980, \pra, 21, 2087

\bibitem[Slattery, Doolen \& DeWitt(1982)]{1982Slattery} Slattery, W.~L., Doolen, G.~D., \& DeWitt, H.~E. \ 1982, \pra, 26, 2255
\bibitem[Suzuki et al.(2011)]{2011Suzu} Suzuki, T., Honma, M., Mao, H., Otsuka, T., \& Kajino, T. \ 2011, \prc, 83, 044619
\bibitem[Suzuki, Nomoto \& Toki(2016)]{2016Suzu} Suzuki, T., Nomoto, K., \& Toki, H. \ 2016, \apj, 817, 163
\bibitem[Timmes \& Woosley(1992)]{1992ApJ...396..649T} Timmes, F.~X. \& Woosley, S.~E.\ 1992, \apj, 396, 649

\bibitem[Toki et al.(2013)]{2013Toki} Toki, H., Suzuki, T., Nomoto, K, Jones, S., \& Hirschi, R. \ 2013, \prc, 88, 015806

\bibitem[Wallace, Woosley \& Weaver(1982)]{1982ApJ...258..696W} Wallace, R.~K., Woosley, S.~E., \& Weaver, T.~A.\ 1982, \apj, 258, 696

\bibitem[Whelan \& Iben(1973)]{1973ApJ...186.1007W} Whelan, J., \& Iben, I.\ 1973, \apj, 186, 1007
\bibitem[Woosley \& Weaver(1994)]{1994ApJ...423..371W} Woosley, S.~E., \& Weaver, T.~A.\ 1994, \apj, 423, 371
\bibitem[Yakovlev \& Shalybkov(1989)]{1989ASPRv...7..311Y} Yakovlev, D.~G. \& Shalybkov, D.~A.\ 1989, \apspr, 7, 311

\bibitem[Yamaguchi et al.(2014)]{2014ApJ...780..136Y} Yamaguchi, H., Eriksen, K.~A., Badenes, C., et al.\ 2014, \apj, 780, 136
\bibitem[Yamaguchi et al.(2015)]{2015ApJ...801L..31Y} Yamaguchi, H., Badenes, C., Foster, A.~R., et al.\ 2015, \apjl, 801, L31
\bibitem[Yang et al.(2018)]{2018ApJ...852...89Y} Yang, Y., Wang, L., Baade, D., et al.\ 2018, \apj, 852, 89

 \end{thebibliography}
\end{document}